\documentclass[twocol]{ametsocV6.1}

\usepackage{rotating}
\usepackage{array}

\newcommand{\plotlayout}{from forecasts initialized with own initial conditions (OIC, left column) and
  operational ECMWF initial conditions (SIC, right column).  Results for physical models
  are shown in the first row (a and d), AIWP models in the second row (b and e), and hybrid
  models in the third row (c and f).  Contributions are color-coded following the legend,
  with ECMWF forecasts identical in both streams and plotted in all panels for reference.
  All calculations are performed on the $0.25^\circ$ WP-MIP common grid.  
  The 95\% confidence interval of the mean is computed with a 1000-member bootstrap and
  shown with semi-transparent shading}

\title{WP-MIP: An Artificial Intelligence, Hybrid, and Physically Based Model Intercomparison Project for Weather Prediction}

\authors{
  Ron McTaggart-Cowan,\aff{a}\correspondingauthor{Ron McTaggart-Cowan, ron.mctaggart-cowan@ec.gc.ca} 
  Linus Magnusson,\aff{b} 
  Inna Polichtchouk,\aff{b} 
  Duncan Ackerley,\aff{c} 
  Martin K\"{o}hler,\aff{d} 
  Barbara Casati,\aff{a}
  Jan-Huey Chen,\aff{e}
  Debra Hudson,\aff{f}
  Masashi Ujiie,\aff{g}
  Nurizana Amir Aziz,\aff{h}
  Massimo Bonavita,\aff{b}
  Zied Ben Bouall\`{e}gue,\aff{b}
  Catherine de Burgh-Day,\aff{f}
  Stephane Chamberland,\aff{a}
  Kyoungmi Cho,\aff{i}
  Caio. A. S. Coelho,\aff{j}
  Rostislav Fadeev,\aff{k}
  Manuel Fuentes,\aff{b}
  Jorge, L. Garcia-Franco,\aff{l}
  Claude Gibert,\aff{c}
  Bruno. S. Guimar\~{a}es,\aff{j}
  Chris Harris,\aff{c}
  Michelle Harrold,\aff{m}
  Syed Husain,\aff{a}
  Molly James,\aff{c}
  Alex Kaltenbaugh,\aff{e}
  Marta Koch,\aff{n}
  Paulo Y. Kubota,\aff{j}
  Eun-Hee Lee,\aff{o}
  Chen Li,\aff{f}
  Wei Li,\aff{p}
  Weiwei Li,\aff{m}
  Lloren\c{c} Lled\'{o},\aff{b}
  Nicholas Loveday,\aff{f}
  Cristian Lussana,\aff{q}
  Zubiar Maalick,\aff{c}
  Mohau J. Mateyisi,\aff{r}
  Amy McGovern,\aff{s}
  Koos van der Merwe,\aff{r}
  Joel Miller,\aff{c}
  Marion Mittermaier,\aff{c}
  Richard Mladek,\aff{b}
  Kathryn Newman,\aff{m}
  Andr\'{e} L. O. Neves,\aff{j}
  John Pill,\aff{c}
  Roland Potthast,\aff{d}
  Maheswar Pradhan,\aff{t}
  Subhrajit Rath,\aff{t}
  David S. Richardson,\aff{b}
  Leo Separovic,\aff{a}
  Michelle Sim\~{o}es Reboita,\aff{u}
  Gregor Skok,\aff{v}
  Ankur Srivastava,\aff{t}
  Mikhail Tolstykh,\aff{k}
  Zhuo Wang,\aff{w}
  Beth J. Woodhams,\aff{c}
  Fanglin Yang,\aff{p}
  Radomir Zaripov,\aff{x}
  Gan Zhang,\aff{w}
  Hongyan Zhu,\aff{f}
}

\affiliation{
  \aff{a}{Environment and Climate Change Canada}\\
  \aff{b}{European Centre for Medium-Range Weather Forecasts}\\
  \aff{c}{United Kingdom Met Office}\\
  \aff{d}{Deutcher Wetterdienst}\\
  \aff{e}{National Oceanic and Atmospheric Administration / Geophysical Fluid Dynamics Laboratory}\\
  \aff{f}{Australian Bureau of Meteorology}\\
  \aff{g}{Japan Meteorological Agency} \\
  \aff{h}{Malaysian Meteorological Department (MET Malaysia)} \\
  \aff{i}{Korea Meteorological Administration} \\
  \aff{j}{Centro de Previs\~{a}o de Tempo e Estudos Clim\'{a}ticos (CPTEC), Instituto Nacional de Pesquisas Espaciais (INPE)} \\
  \aff{k}{Marchuk Institute of Numerical Mathematics Russian Academy of Sciences (INM RAS), and Hydrometcentre of Russia} \\
  \aff{l}{Escuela Nacional de Ciencias de la Tierra, UNAM, Mexico} \\
  \aff{m}{National Science Foundation National Center for Atmospheric Research (NSF NCAR)} \\
  \aff{n}{Imperial College London} \\
  \aff{o}{Korea Institute of Atmospheric Prediction Systems} \\
  \aff{p}{National Oceanic and Atmospheric Administration Office of Modeling and Development} \\
  \aff{q}{Norwegian Meteorological Institute} \\
  \aff{r}{Council for Scientific and Industrial Research} \\
  \aff{s}{University of Oklahoma and Brightband} \\
  \aff{t}{Indian Institute of Tropical Meteorology} \\
  \aff{u}{Universidade Federal de Itajub\'{a} (UNIFEI), Brazil} \\
  \aff{v}{Faculty of Mathematics and Physics, University of Ljubljana} \\
  \aff{w}{University of Illinois Urbana-Champaign} \\
  \aff{x}{Hydrometcentre of Russia}
}

\abstract{Rapid progress in the field of machine-learning for weather prediction has led to the emergence of algorithms whose forecasting skill can exceed that of traditional physically based models.  This development represents an opportunity to improve the quality of forecasting services provided by operational centers, particularly given the speed at which machine-learning based models generate predictions.  Despite the clear promise of these systems, questions remain about the ability of the current generation of machine-learning models to generate physically consistent predictions of the full suite of required forecast fields under all conditions.  Answering these questions will require careful comparisons between the well-understood physically based models, current state-of-the-art machine-learning models, and the hybrid models that combine elements of these two archetypes.  The Weather Prediction Model Intercomparison Project (WP-MIP) is a World Meteorological Organization-supported initiative whose initial goal is to create a centralized database of physically based, machine-learning, and hybrid model forecasts to enable a distributed assessment and evaluation effort.  The first instance of WP-MIP focuses on global deterministic predictions using both center-specific and common initializations to facilitate sensitivity studies.  Forecasts contributed by institutions across six continents will be used to develop AI-ready verification techniques that highlight the strengths and weaknesses of each class of prediction system, with the goal of establishing best-practice guidance to model developers and national weather centers.  The broad engagement of the operational and forecast-evaluation communities in WP-MIP will ensure that the project’s results are highly relevant to the development and deployment of next-generation weather prediction systems.}

\begin{document}

\maketitle

%
%
%
\statement
The World Meteorological Organization’s Weather Prediction Model Intercomparison Project aims to identify the strengths
and weaknesses of weather forecasts generated by numerical models based on artificial intelligence, physical principles,
and hybrids between the two.  Global predictions from models under development at participating national meteorological
and hydrological services are compared within a well-controlled experimental framework.  Although the artificial
intelligence-based systems generate highly accurate 10-day forecasts, they suffer from unphysical smoothing.  Hybrids
that combine artificial intelligence-based guides with physically based models benefit from synergies between their
components.  The project research plan builds on these findings to develop verification strategies that reliably
assess forecast skill across modeling paradigms in support of planning and decision-making for next-generation
operational numerical weather prediction systems.
%
\capsule
The WMO Weather Prediction Model Intercomparison Project combines global forecasts generated by artificial-intelligence, physical,
and hybrid models from operational centres around the world within a common experimental protocol.  The resulting
archive will be used to develop new verification techniques designed to assess the strengths and
weaknesses of the different modeling paradigms. These innovations will inform the development of WMO guidance for
national meteorological and hydrological services as they prepare to operationalize the next generation of numerical
modeling systems.

%

%

\section{Introduction}
\label{intro}
National meteorological and hydrological services are responsible for providing vital environmental predictions to
users around the world \citep{WMO17}.  These forecasts are essential for economic growth and for the preservation
of lives and property across sectors as diverse as public safety, agriculture, resource management, power generation, and transportation.
The United Nations' Early Warnings for All (EW4All) initiative aims to ensure that this critical information is globally
accessible by the end of 2027 \citep{WMO22}.

This ambitious timeline overlaps with the rapid emergence of artificial intelligence (AI)-based weather prediction
(AIWP) models that can generate forecasts whose skill is comparable to that of leading operational physically
based models for a small fraction of the computational cost \citep{Rabier26}.
Although the first generation of deterministic AIWP systems \citep{Keisler22, Pathak22, Bonev23, Bi23, Lam23, Lang24}
suffered from problems including excessive smoothing, poor dynamical balance, and unphysical results 
\citep{Bonavita24}, focused development efforts have mitigated these weaknesses \citep{Kochkov24,Bodnar25,Bonev25,Moldovan25,Sha25,Lang26}.
The ``quiet revolution'' of physically based NWP \citep{Bauer15} has undergone a radical transformation
as private industry and academic partners join the effort to improve forecasts for a wide variety of end users \citep{Shenolikar25}.

Despite the many successes of AIWP \citep{Hakim24}, the current generation of algorithms cannot directly
replace operational physical models because they produce only a subset of the required forecast fields.
They also suffer from systematic errors in prediction of high-impact weather \citep{Radford25b,Davis26}, including
a weak-intensity bias for tropical cyclones \citep{DeMaria25}.  In response to such concerns \citep{Smith26}, hybrid models
have emerged that combine AI and physically based approaches.  Although these archetypes could be combined in
many ways, spectral nudging of a physical model to AIWP large-scale predictions is currently the primary
hybridization strategy \citep{Husain25,Polichtchouk25,Polichtchouk26}.  The resulting
predictions benefit from skillful medium-range AIWP guidance while generating a complete set
of physically consistent fields with the full range of local variability and extremes.

These new modeling paradigms present both opportunities and challenges to national meteorological
and hydrological services.  Rapid AIWP forecasts can help to ``democratize'' forecast production \citep{Khadir25}, potentially
fulfilling EW4All objectives; however, robust guarantees of prediction quality are required to support
essential forecasting and warning services.  The forecast verification community is therefore actively engaged
in augmenting current practices and developing new ``AI-ready'' evaluation techniques to assess
attributes of model predictions that are generally -- perhaps unjustifiably -- taken for granted in physical-model forecasts.
These verification development activities are being promoted by the WMO Integrated Processing and
Prediction System (WIPPS), the framework that governs the operational exchange of forecast products, international
evaluation standards, and associated guidance to support WMO Members on topics including AIWP \citep{WMO23}.

In response to this superposition of scientific and operational objectives, the WMO Working Group on Numerical
Experimentation (WGNE) initiated the Weather Prediction Model Intercomparison Project (WP-MIP)
in late 2024 to assess the strengths and weaknesses of AIWP, hybrid, and physical model predictions \citep{wpmip25}.
The first phase of WP-MIP saw the creation of an extensive archive of global guidance from models being
developed to support operational forecasting. The project's focus has now shifted to
verification research efforts, coordinated by members of the WMO Joint Working Group on
Forecast Verification Research.

The WP-MIP objectives complement those of AI-focused projects such as WeatherBench 2 \citep{Rasp24} and
\citet{Radford25a}; however, its design is closer to that of flagship WMO initiatives like the Coupled Model
Intercomparison Project \citep{Durack25} than it is to standard AI benchmarking efforts.
Recognizing that diversity is essential to the success of such intercomparisons,
global engagement has been a cornerstone of WP-MIP.  Inclusion of all current modeling paradigms
has maximized the breadth of contributions, enabling robust assessments within and across
the different classes of models.  The active involvement of verification experts
has ensured that WP-MIP forecasts can be easily integrated into existing workflows, contain a
rich set of predicted fields to support verification research efforts, and fit into
an extensible framework to accommodate community-initiated topical subprojects.
This commitment to engagement fosters capacity-building across
National Meteorological and Hydrological Services -- 17 of which are involved with the project -- in
support of the EW4All initiative. Taken together, these attributes help WP-MIP to fulfill both
the ``calls to action'' of \citet{Panago24} and the
\citet{Brocker26} request for operationally relevant benchmarking, while the project's designation as a
WIPPS AI Pilot Project \citep{WIPPS26} ensures that it has a direct impact on AI integration
into next-generation operational NWP systems.

The WP-MIP protocol and descriptions of participating models are presented in section~\ref{data}.
Preliminary results are shown in section~\ref{results}, followed by a description of ongoing verification
efforts (section~\ref{verif}).  The study concludes with a discussion of the next steps for the project
in section~\ref{discussion}.

\section{Project Protocol and Models}
\label{data}
\begin{figure}[t]
  \centering
  \includegraphics[width=\columnwidth]{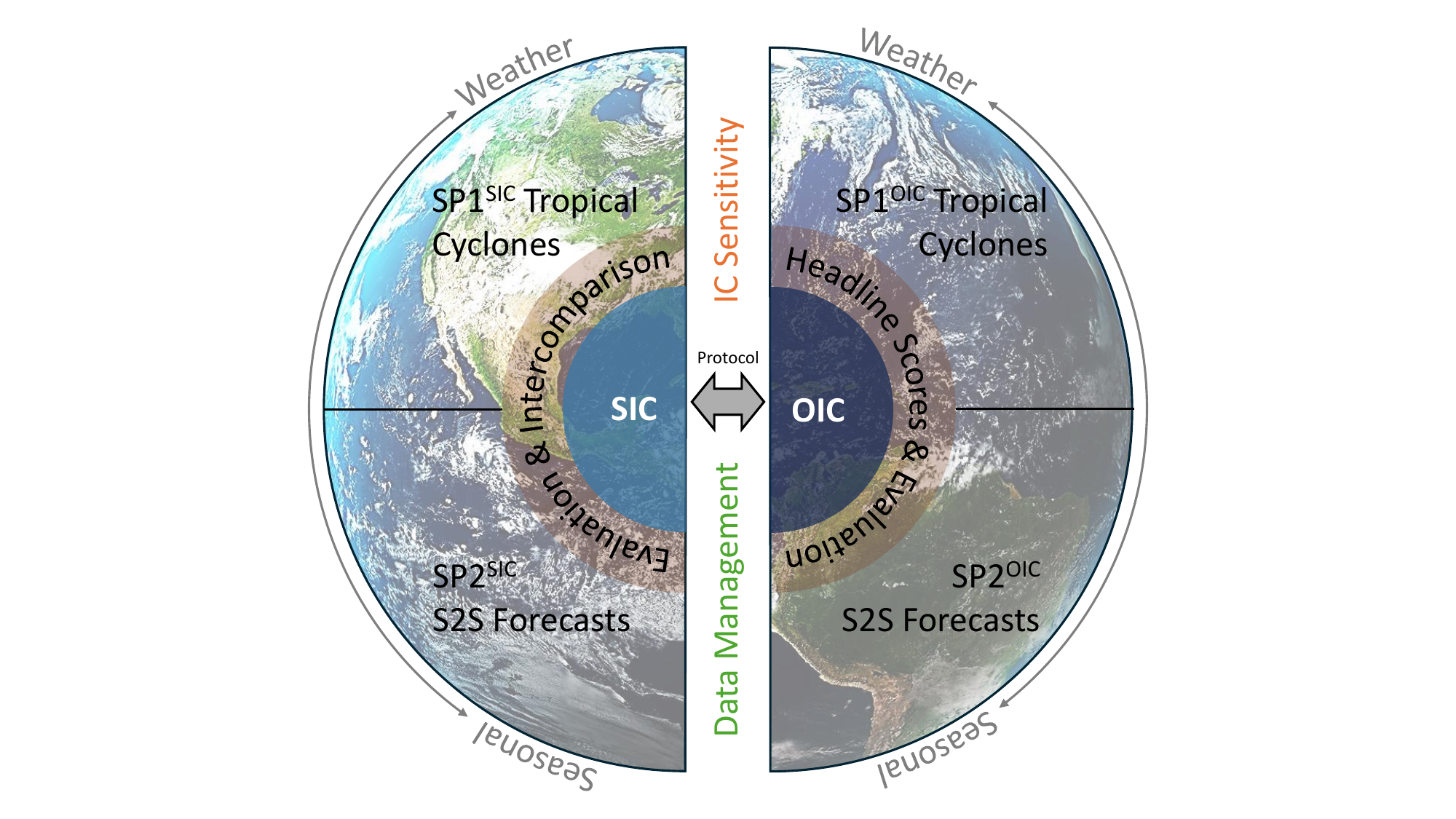}
  \caption{Project overview schematic.  The core protocol (blue) is divided by initial condition (IC)
    specifications into the ``same initial condition'' (SIC)
    and ``own initial condition'' (OIC) streams, described in more detail in section~\ref{data}\ref{protocol}.  The
    two WP-MIP subprojects (SP1 and SP2) are described in sections~\ref{data}\ref{protocol}\ref{sp1} and
    \ref{data}\ref{protocol}\ref{sp2}, respectively.}
  \label{earth}
\end{figure}

The condensed description of WP-MIP provided here is supported by the project overview
schematic (Fig.~\ref{earth}). Details can be found in the project white paper \citep{wpmip25}.

\subsection{Experimental Protocol}
\label{protocol}
The WP-MIP protocol centers on global deterministic 10-day forecasts for 2024, providing
robust sampling of day-to-day weather.  A total of 122 forecasts are initialized at 3-day
intervals to ensure functional independence at the synoptic scales.  A longer period would have
extended sampling to include lower-frequency modes and increased the population of
extreme events.  However, many AIWP training regimens extend through 2022, with fine-tuning
periods appended thereafter.  Avoiding these training periods is essential to ensuring the
robustness of AIWP and hybrid model evaluations.

Predictions of 3D state variables and key single-level fields are remapped to a $0.25^\circ$ latitude-longitude
grid at 6-hourly intervals.  This common grid resembles that of most AIWP models, balancing
storage and transfer limitations with the need to quantify extremes and to evaluate forecast skill
over a broad range of scales.  The protocol is flexible enough to encourage contributions from AIWP
models that can provide only a subset of the requested fields; however, maximum compliance is
encouraged to facilitate studies
of extreme weather (e.g., near-surface fields and precipitation), energy budgets (e.g., accumulated radiative and
turbulent fluxes), clouds (e.g., cloud cover over distinct layers), and polar processes (e.g., sea ice properties).

\begin{figure}[t]
  \centering
  \includegraphics[width=\columnwidth]{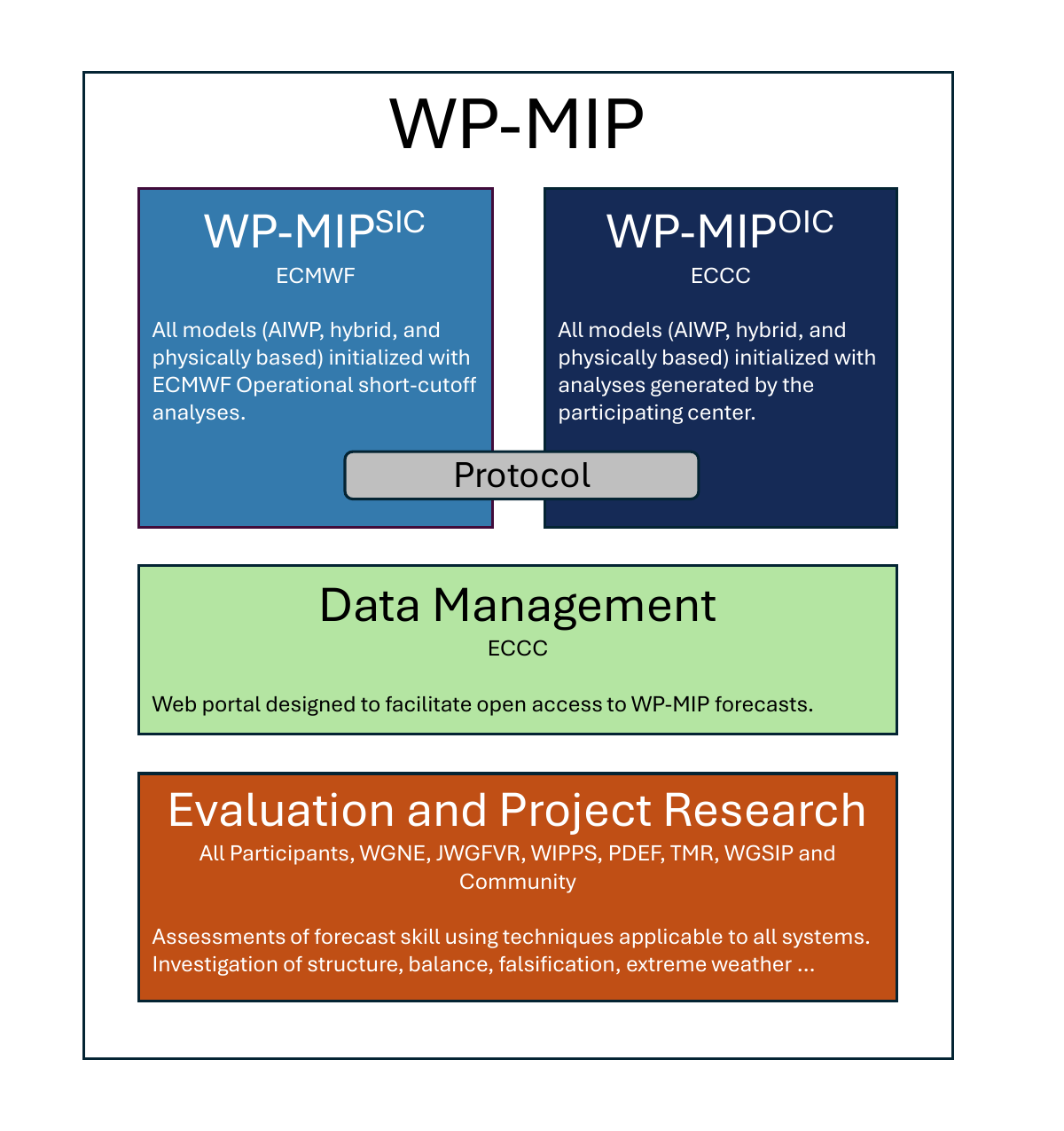}
  \caption{Functional structure of WP-MIP.  Leadership of each component of the project is identified
    below the component title using institute names (Table~\ref{centre_table}) and WMO acronyms for the Working
    Group on Numerical Experimentation (WGNE), the Joint Working Group on Forecast Verification Research
    (JWGFVR), the WMO Integrated Processing and Prediction System (WIPPS), the Working Group on Predictability Dynamics
    and Ensemble Forecasting (PDEF), the Working Group on Tropical Meteorology Research
    (TMR) and the Working Group on Seasonal to Interdecadal Prediction (WGSIP).}
  \label{structure}
\end{figure}

The WP-MIP protocol defines two distinct forecast streams (Figs.~\ref{earth} and \ref{structure}). In the
operationally based ``own initial condition'' (OIC) stream, forecasts are initialized with each center's own analyses.
The ``same initial condition'' (SIC) stream requires initialization from operational ECMWF analyses,
a design that resembles that of the DIfferent MOdels Same Initial Conditions project \citep{Magnusson22}.
Intra-stream comparison provides information about the relative
skill between modeling systems, while cross-stream comparison quantifies initial-condition sensitivity, a topic that
remains largely unstudied for AIWP and hybrid models.

The global deterministic focus of WP-MIP encourages worldwide participation while limiting data volume.
Although medium-range predictions are better considered probabilistically \citep{Mylne25},
including ensembles would have required a trade-off with common-grid resolution.
Plans to expand TIGGE \citep{Swinbank16} to include AIWP and hybrid models represent an extension of
WP-MIP that will accelerate future ensemble-based intercomparison work. 

\subsubsection{Subproject 1: Tropical Cyclones}
\label{sp1}
The 3-day initialization interval prescribed in the core protocol provides limited sampling
of infrequent events like tropical cyclones \citep{Chen23}.  Subproject 1 (SP1; Fig.~\ref{earth}) is an
optional extension that asks contributors to provide tropical cyclone tracking inputs from daily forecast initializations.
Algorithms developed by JMA \citep{Yamaguchi17} and GFDL \citep{Marchok21} have been used to create
a common set of WP-MIP tracking files.
These data will facilitate tropical cyclone evaluation by reducing sampling and tracking uncertainties.
The SP1 leads (coauthors J.-H. Chen and M. Ujjie) are coordinating with the WMO Working Group on
Tropical Meteorology Research to ensure that efficient tropical cyclone diagnostic and verification
activities play an important role in model evaluation.

\subsubsection{Subproject 2: Extended-Range Prediction}
\label{sp2}
The focus of current AIWP and hybrid models is on medium-range forecasting; however,
the WMO Working Group on Seasonal to Interannual Prediction has prioritized assessment of
their potential for extended-range prediction.
Subproject 2 (SP2; Fig.~\ref{earth}), led by coauthor D. Hudson, consists of an optional extension to
15~d forecasts that will enable preliminary evaluations
of S2S-relevant features including the MJO, climate indices,
monsoons, and convectively coupled tropical waves.
The deterministic WP-MIP framework is likely suboptimal for quantitative S2S analyses,
while the use of a single year means that the influence of low-frequency variability
on forecast skill cannot be assessed.  However, the initial guidance derived from the
lightweight SP2 extension will be used to refine protocols for future intercomparison
projects tailored to S2S evaluations.

\subsection{Participating Models}
\label{models}
As of writing, 11 institutions have contributed forecasts to WP-MIP (Table~\ref{centre_table}).
Future submissions from academic and private partners would be a welcome expansion
of the project's scope. A synthesis of key model attributes is presented here,
with more detailed information available on the WP-MIP model description hub \citep{WPMIPdescrip26}.  

\begin{table*}[t]
  \small
  \begin{center}
    \begin{tabular}{lp{3in}p{2in}}
      \hline\hline
      Center Acronym & Center Name & Country/Region \\
      \hline\hline
      CPTEC/INPE & Centro de Previs\~{a}o do Tempo e Estudos Clim\'{a}ticos / National Institute for Space Research & Federative Republic of Brazil \\
      CMA & China Meteorological Administration & People's Republic of China \\
      ECCC & Environment and Climate Change Canada & Canada \\
      ECMWF & European Centre for Medium-Range Weather Forecasts & Europe \\
      GFDL & Geophysical Fluid Dynamics Laboratory & United States of America \\
      DWD & Deutscher Wetterdienst & Federal Republic of Germany \\
      JMA & Japan Meteorological Agency & Japan \\
      KMA/KIAPS & Korea Meteorological Administration / Korea Institute of Atmospheric Prediction Systems & Republic of Korea \\
      UKMO & Met Office & United Kingdom of Great Britain and Northern Ireland \\
      NOAA & National Oceanographic and Atmospheric Administration & United States of America \\
      RAS & Russian Academy of Sciences & Russian Federation \\
      \hline
    \end{tabular}
  \end{center}
  \caption{Institutions that have contributed forecasts to WP-MIP.}
  \label{centre_table}
\end{table*}
The long history of physical model development has led to broad diversity in model
designs (Table~\ref{pm_table}).  Conversely, most of the AIWP systems being developed by
operational centers rely on graph neural network architectures
(Table~\ref{ai_table}).  Although hybrid models inherit structural diversity from their
constituents, current hybridization strategies focus uniquely on spectral
nudging to AIWP guidance (Table~\ref{hy_table}).

\begin{sidewaystable*}
  \small
  \begin{center}
    \begin{tabular}{p{0.7in}p{2.5in}>{\centering}p{0.65in}cc>{\centering}p{0.65in}p{2in}}
      \hline\hline
      Center & Description & Grid & Levels & Top & Step & References \\
      \hline
      CPTEC/INPE & Spectral dynamical core with semi-implicit semi-Lagrangian advection & $0.18^\circ$ & 64 & 0.016~hPa & 200~s & \citet{Coelho22,Guimaraes19} \\
      DWD & Finite-volume dynamical core on an icosahedral grid & $0.12^\circ$ & 120 & 75~km & 120~s & \citet{Zangl15,Muller25} \\
      ECCC & Iteratively implicit semi-Lagrangian on a Yin-Yang grid & $0.135^\circ$ & 84 & 0.1~hPa & 450~s & \citet{Girard14,McTaggart-Cowan19b} \\
      ECMWF & Semi-implicit, semi-Lagrangian spectral dynamical core & $0.08^\circ$ & 137 & 0.01~hPa & 450~s & \citet{Malardel16,ECMWF23} \\
      GFDL & Finite-volume dynamical core on a cubed sphere grid & $0.06^\circ$ & 91 & 40~hPa & 150~s & \citet{Lin04,Zhou22} \\
      JMA & Semi-implicit, semi-Lagrangian spectral dynamical core & $0.125^\circ$ & 128 & 0.01~hPa & 300~s & \citet{jma25} \\
      KMA/KIAPS & Spectral element model on a cubed-sphere grid & $0.125^\circ$ & 91 & 0.01~hPa & 25~s & \citet{Hong18} \\
      NOAA & Finite-volume dynamical core on a cubed-sphere grid & $0.12^\circ$ & 127 & 80~km & 150~s & \citet{Yang20} \\
      RAS & Semi-implicit, semi-Lagrangian dynamical core on a latitude-longitude grid & $0.08^\circ\,\text{--}\,0.13^\circ$ (lat) $\times\ 0.1^\circ$ (lon)& 104 & 0.05~hPa & 270~s & \citet{Tolstykh17,Tolstykh24} \\
      UKMO & Semi-implicit, semi-Lagrangian dynamical core on a latitude-longitude grid & $0.140625^\circ \times 0.09375^\circ$ & 70 & 80~km & 240~s & \citet{Walters19,Guiavarch25,Willet25} \\
      \hline
    \end{tabular}
  \end{center}
  \caption{Description of physical models participating in WP-MIP.  Nominal grid spacings are approximate for unstructured grids and spectral models.}
  \label{pm_table}
\end{sidewaystable*}

\begin{sidewaystable*}
  \small
  \begin{center}
    \begin{tabular}{lp{2in}c>{\centering}p{1in}>{\centering}p{0.7in}cccl}
      \hline\hline
      Center & Description & Training & Fine Tuning & Loss & Grid & Levels & Step & References \\
      \hline
      CMA & Encode-process-decode architecture with multi-scale encoder/decoder design and learnable latent queries & ERA5 & None & MSE & $0.25^\circ$ & 13 & 6~h & \\
      ECCC & Graph neural network based on \citet{Lam23} & ERA5 & ECMWF analyses & MSE & $0.25^\circ$ & 17 & 6~h & \citet{Subich26} \\
      ECMWF & Graph neural network combined with a sliding-window transformer & ERA5 & ECMWF analyses & MSE & $0.25^\circ$ & 13 & 6~h & \citet{Lang24} \\
      DWD & Graph neural network on the DWD physical-model grid with attention mechanism & ICON-DREAM & None & MSE & 13~km & 13 & 3~h & \\
      JMA & Graph neural network based on \citet{Lam23} & ERA5 & ECMWF analyses & MSE & $0.25^\circ$ & 13 & 6~h & \\
      NOAA & Graph neural network based on \citet{Lam23} & ERA5 & ECMWF analyses, NOAA analyses & MSE & $0.25^\circ$ & 13 & 6~h & \citet{Tabas25} \\
      \hline
    \end{tabular}
  \end{center}
  \caption{Description of AIWP models participating in WP-MIP.}
  \label{ai_table}
\end{sidewaystable*}

\begin{sidewaystable*}
  \small
  \begin{center}
    \begin{tabular}{lp{2in}>{\centering}p{1in}>{\centering}p{1.2in}>{\centering}p{0.65in}>{\centering}p{0.65in}p{1.3in}}
      \hline\hline
      Center & Description & Physical Model Configuration & AIWP Configuration & Cutoff Wavelength & Relaxation Time & References \\
      \hline
      ECCC & Spectral nudging of the ECCC physical model to the ECCC AIWP solution & Table~\ref{pm_table} & Table~\ref{ai_table} & 2500~km & 12~h & \citet{Husain25} \\
      ECMWF & Spectral nudging of the ECMWF physical model to a model-level form of the ECMWF AIWP solution & Table\ref{pm_table} & Similar to table~\ref{ai_table}, but on a $1^\circ$ grid with vertical levels as in table~\ref{pm_table} & 2000~km & 12~h & \citet{Polichtchouk26} \\
      UKMO & Spectral nudging of the UKMO physical model to a model-level form of the ECMWF AIWP solution & Table~\ref{pm_table} & Similar to ECMWF entry in table~\ref{ai_table}, but on a $1^\circ$ grid with vertical levels as in table~\ref{pm_table} & 2000~km & 6~h & \\
      \hline
    \end{tabular}
  \end{center}
  \caption{Description of hybrid models participating in WP-MIP.}
  \label{hy_table}
\end{sidewaystable*}

\subsection{Reference Datasets for Verification}
\label{analyses}
Analyses from five contributing centers (ECCC, ECMWF, NOAA, UKMO and RAS) are available
on the WP-MIP common grid to mitigate the problem of incestuous ``own analysis'' evaluation,
in which systematic errors in the data assimilation cycle favor forecasts from the model
native to the system \citep{Caron25}. 
Other regridded products will be incorporated into the WP-MIP archive when shared
access is likely to improve research efficiency. For example, GPM IMERG precipitation
estimates \citep{Huffman23} conservatively remapped to the WP-MIP
grid will be included as a reference dataset for evaluation teams.

\subsection{Project Data and Software}
\label{license}
All WP-MIP participants follow FAIR principles (fair, accessible,
interoperable, and reusable), for sharing of project data and software to the greatest
extent possible. This goal is supported by the application of the Creative Commons CC-BY-4.0
license to all project outputs (see Availability Statement).
The WP-MIP data portal \citep{WPMIPportal26} is accessible directly or through the
project website \citep{WPMIPhome25} and contains up-to-date information about data
availability.

As of July 2026, the WP-MIP archive contains 42 forecast contributions and
5 analysis datasets totalling 31~TB of data.  New contributions can be added,
but existing forecasts will not be modified.  The data are stored in WMO-standard GRIB2
format in monthly files split by variable and level, minimizing the volume of transfers
required for verification tasks while facilitating integration into operational verification
workflows.  An open issue tracking system is used to monitor tasks and problems related
to WP-MIP data \citep{WPMIPgithubdata26}, including an effort to enhance accessibility
by supporting data access through the cloud-native Zarr interface.  Archive management software is made
available and tracked in a parallel repository \citep{WPMIPgithubsoftware26}.

\section{Preliminary Results}
\label{results}
Despite WP-MIP's focus on AI-ready verification techniques, baseline assessments
of standard scores (section~\ref{results}\ref{scores}) and spatial variability
(section~\ref{results}\ref{spectra}) are needed to place project forecasts in the
context of international standards and previous studies.

\subsection{Headline Scores}
\label{scores}
\begin{figure*}[t]
  \centering
  \includegraphics[width=\textwidth]{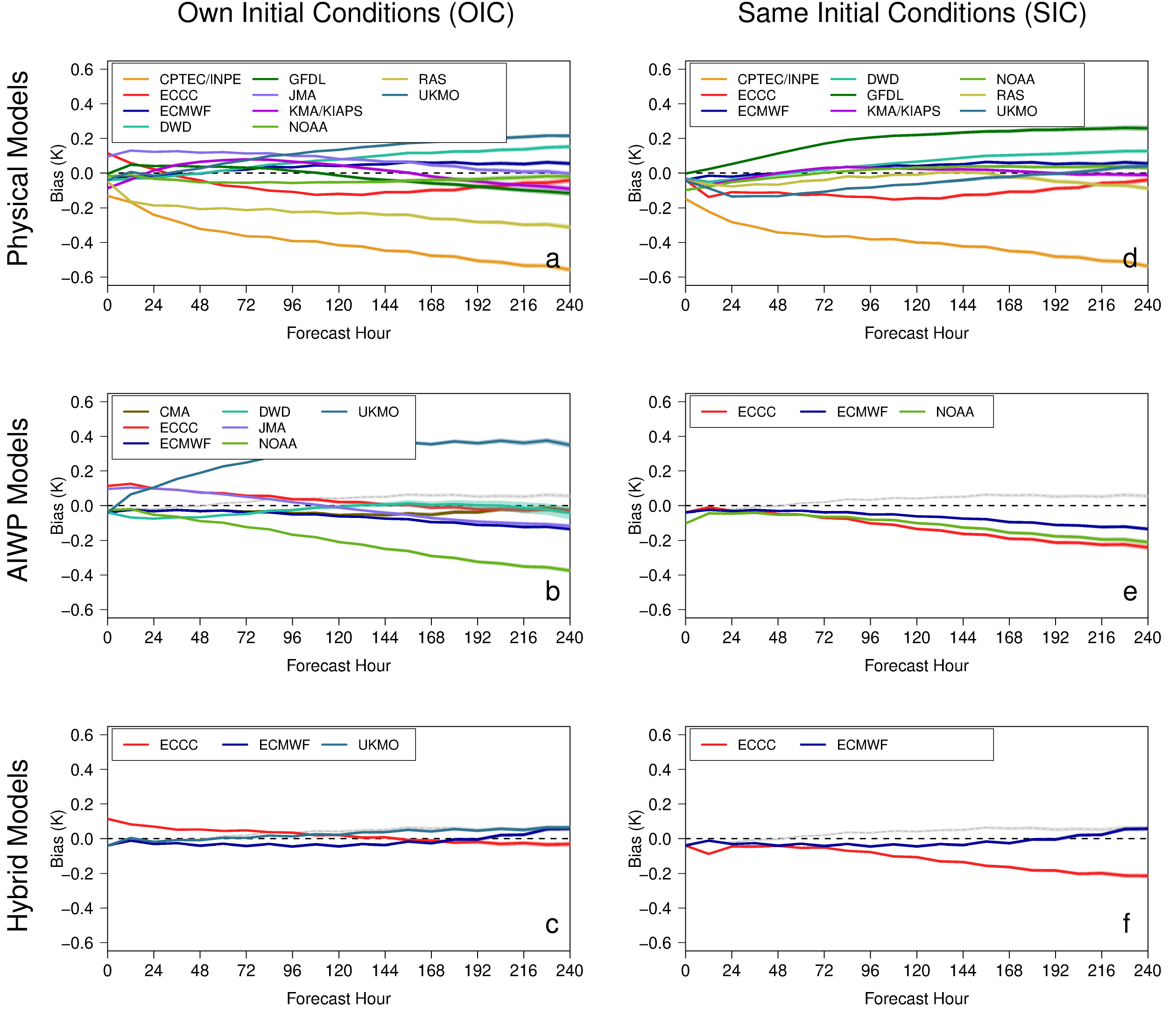}
  \caption{Bias (mean of forecast minus analysis) of global 500~hPa temperature with respect to
    reference analyses (section~\ref{data}\ref{analyses}) \plotlayout.  Biases are
    computed independently against each analysis and contribute equally to the sampling distribution.
    The ECMWF physical-model biases are plotted as a grey dashed
    line in the second and third rows, and the 0-K bias is denoted with a
    dashed black line in all panels for reference.}
  \label{bias}
\end{figure*}

\begin{figure*}[t]
  \centering
  \includegraphics[width=\textwidth]{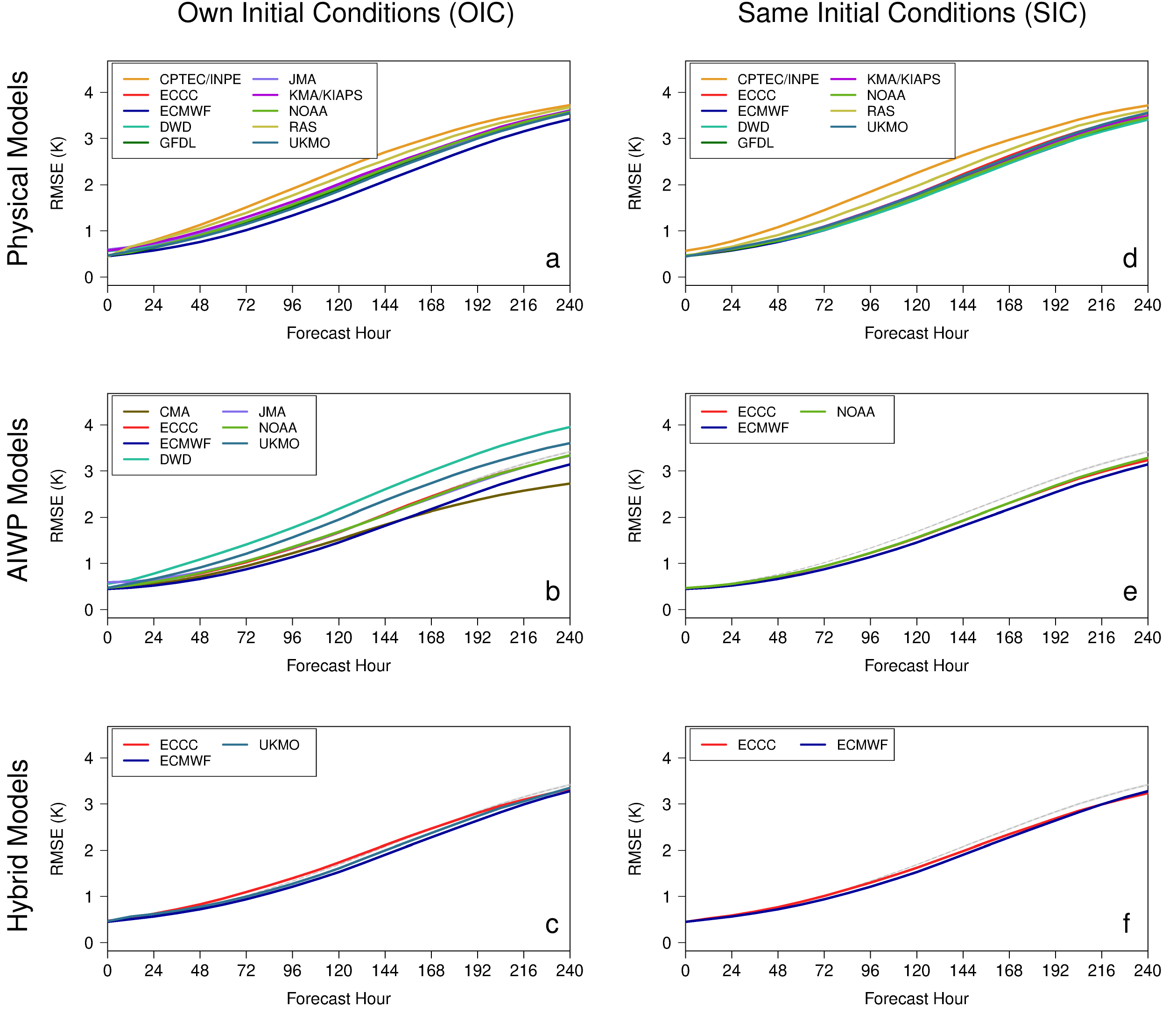}
  \caption{Root mean square error of global 500~hPa temperature with respect to
    reference analyses (section~\ref{data}\ref{analyses}) \plotlayout.  Errors are
    computed independently against each analysis and contribute equally to the sampling distribution.
    The ECMWF physical-model RMSEs are plotted as a grey dashed
    line in the second and third rows for reference.}
  \label{rmse}
\end{figure*}

\begin{figure*}[t]
  \centering
  \includegraphics[width=\textwidth]{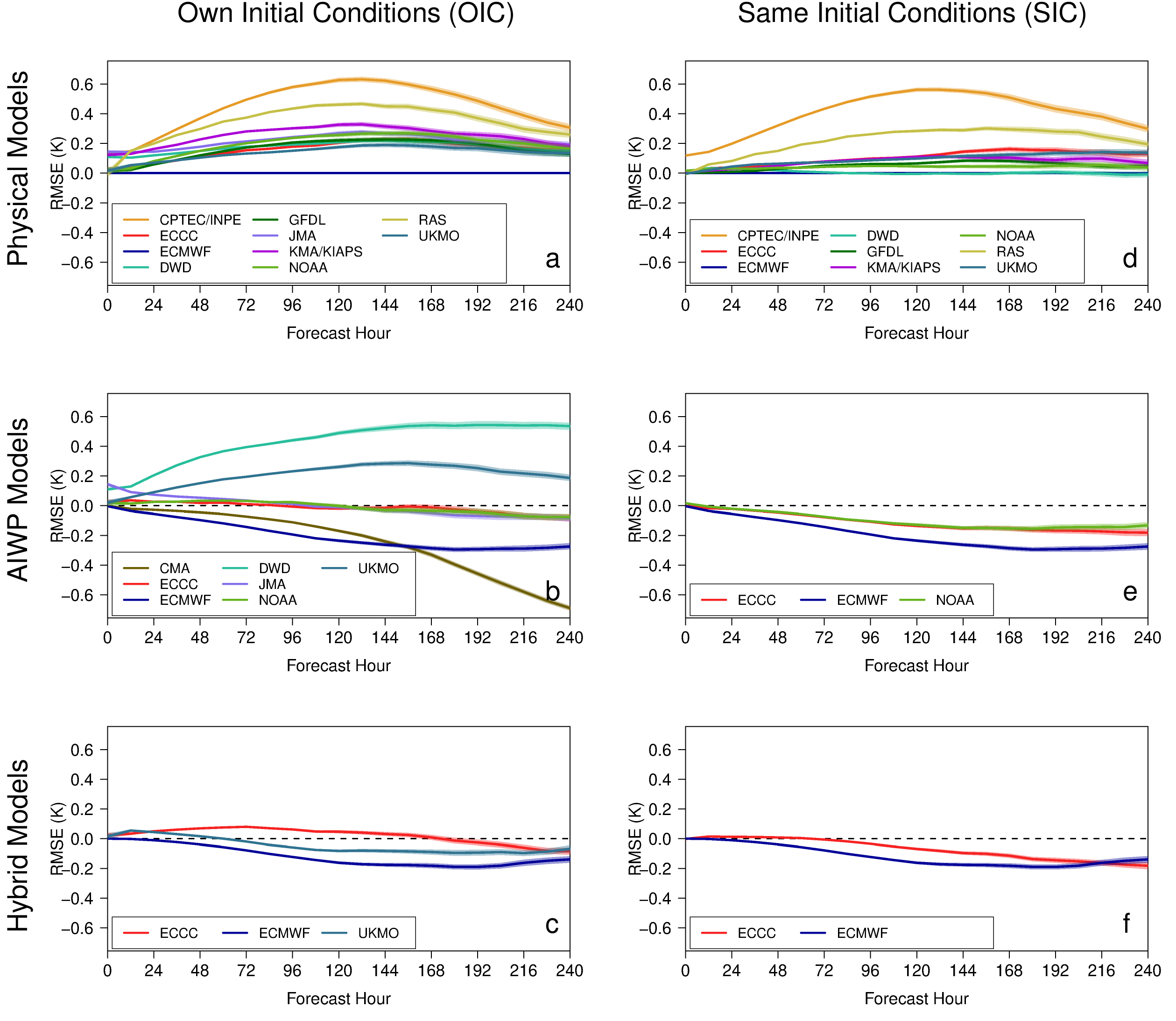}
  \caption{Departure of global 500~hPa temperature RMSE from that of the ECMWF physical-model forecast
    (forecast RMSE minus ECMWF physical-model forecast RMSE),
    computed with respect to reference analyses (section~\ref{data}\ref{analyses}) \plotlayout.
    Departures are computed independently against each analysis and contribute equally to the sampling
    distribution.}
  \label{rmse_rel}
\end{figure*}

The RMSE is a staple for forecast evaluation because of
its simple formulation and straightforward interpretation.  However, the use of
this ``L2'' or Euclidian norm in the loss functions used to train many
first-generation AIWP models means that these algorithms focus almost exclusively
on reducing RMSE. This appears to offer a significant but potentially
superficial advantage to AIWP and hybrid models.  The reliance on
RMSE also encourages smoothing in AIWP forecasts as the network learns
to reduce loss by eliminating poorly predicted smaller-scale features, thereby
minimizing the double-penalty errors induced by feature misplacement \citep{Hoffman95,Ebert13}.
Despite these limitations, assessment of the bias and RMSE of 500~hPa temperature
(Figs.~\ref{bias}\textendash\ref{rmse_rel}) provides information about mass-field
predictions while avoiding the complexity of levels that regularly intersect the
planetary boundary layer.

Most physical models maintain
a steady mid-tropospheric temperature bias throughout their integrations (Fig.~\ref{bias}a and d);
however, CPTEC/INPE forecasts cool substantially with lead time, UKMO and DWD forecasts warm slightly,
and ECCC predictions suffer from transient cooling.  The sources of these systematic
errors are currently under investigation by the contributing centers.  Also unexpected is a long
initialization memory in GFDL predictions: a near-neutral bias in OIC-stream
forecasts (Fig.~\ref{bias}a) is replaced by a long-lived warm bias in SIC integrations
(Fig.~\ref{bias}d).

Although the amplitudes of temperature biases in AIWP models are generally similar to their
physically based counterparts, most contributions exhibit a cooling trend throughout the
forecast period (Fig.~\ref{bias}b and e).  This may be the result of systematic
temperature differences between ERA5 and the analyses used for
fine-tuning \citep{Bouallegue24}.  As the exception to ERA5-based training regimens
(Table~\ref{ai_table}), DWD AIWP
forecasts exhibit a stable temperature bias, potentially
providing guidance for revisions to training and fine-tuning protocols.

The limited number of hybrid models makes it difficult to draw general
conclusions (Fig.~\ref{bias}c and f); however, the ECCC hybrid system clearly inherits the cooling
trend of its AIWP guide (compare Figs.~\ref{bias}e and f).  A similar comparison between the
UKMO AIWP and hybrid-model biases is not possible because the latter uses
an earlier version of the ECMWF AIWP model as a guide (Table~\ref{hy_table}).  The change in sign between
the ECMWF AIWP and hybrid-model biases (compare Fig.~\ref{bias}b and c) likely
arises from different AIWP configurations
(Tables~\ref{ai_table} and \ref{hy_table}), rather than the spectral nudging procedure itself.

The RMSE-estimated accuracy of physical-model contributions to the OIC stream aligns with
expectations: ECMWF forecasts outperform a cluster of models from
other centers (Figs.~\ref{rmse}a and \ref{rmse_rel}a).  Consistent
with DIMOSIC results \citep{Magnusson22}, all physical models benefit from
initialization with ECMWF analyses in the SIC stream, with the RMSE of the
DWD model becoming indistinguishable from that of the ECMWF system itself
(Fig.~\ref{rmse}d and \ref{rmse_rel}d).

Despite their differing OIC-stream initializations, all of the AIWP models that have directly
adopted the \citet{Lam23} system produce forecasts whose RMSEs resemble
those of the ECMWF physical model (Figs.~\ref{rmse}b and \ref{rmse_rel}b).
Architecture enhancements in the ECMWF AIWP model have led to further error
reductions even within the graph neural network context.  Accelerated
error growth in the DWD AIWP model is the result of
a training regimen that optimizes 3-h forecasts, without the autoregressive
rollouts employed by other models (not shown).  Forecasts from the
CMA model have the lowest RMSE values beyond day 7; however, it will be shown in
section~\ref{results}\ref{spectra} that this is achieved through suppression of
unpredictable scales.

A question that surrounds AIWP is how these models represent
growth of initial-state errors \citep{Selz23}.  Day 3\textendash7 RMSE reductions in SIC-stream ECCC and NOAA AIWP predictions
relative to their OIC-stream counterparts (Fig.~\ref{rmse_rel}b and e) suggest that these models are able
to extract information from high-quality initializations.
However, this conclusion is tempered by the use of ERA5 and ECMWF
analyses in the ECCC and NOAA training regimens, a strategy that may leave
``memory'' within the AIWP model that favors ECMWF initializations.
If the anticipated, independently trained SIC-stream DWD AIWP contribution
(Table~\ref{ai_table}) fails to outperform its OIC-stream counterpart,
additional ECCC and NOAA fine-tuning experiments will be needed to
identify regimens that generate models that are sensitive
to the quality of -- rather than a memory of -- the initializing analyses.

The ECCC and UKMO hybrid models benefit substantially from their guiding AIWP predictions, with
RMSE values that closely match those of the ECMWF physical model reference
(Fig.~\ref{rmse}c and \ref{rmse_rel}c).  A near-constant 0.1~K to 0.2~K
RMSE deterioration between each hybrid model and its AIWP guide
(Fig.\ref{rmse}b and c) appears be the result of increased mesoscale variability,
a subject to be investigated in more detail hereafter.

\subsection{Spectral Decomposition}
\label{spectra}
The importance of scale-dependent verification has increased with the advent
of AIWP models that may inflate variability in the more-predictable large
scales while suppressing smaller-scale features to avoid double penalties \citep{Casati25}.
The power spectra of predictions generated by such a model would not be consistent with
the observed atmosphere, a form of ``falsification'' that diminishes the system's
credibility (section~\ref{verif}\ref{tech}\ref{falsification}).

\begin{figure*}[t]
  \centering
  \includegraphics[width=\textwidth]{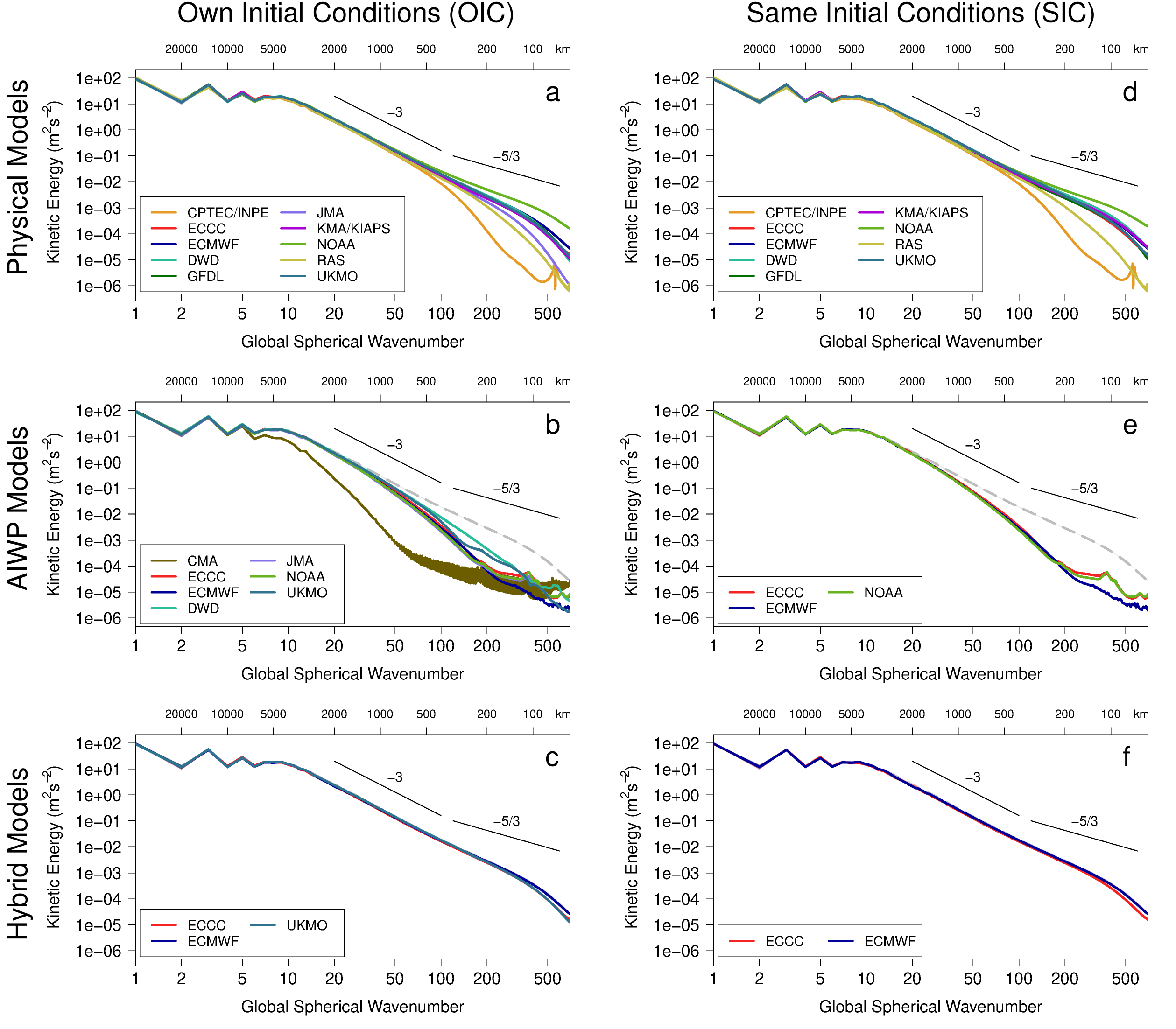}
  \caption{Spectral kinetic energy density for global 250~hPa winds at day 10 \plotlayout.  The -3 and
    $-\frac{5}{3}$ slopes are plotted with thin black lines, and the energy spectrum
    of the ECMWF analysis with a grey dashed line for reference.}
  \label{spec}
\end{figure*}

\begin{figure*}[t]
  \centering
  \includegraphics[width=\textwidth]{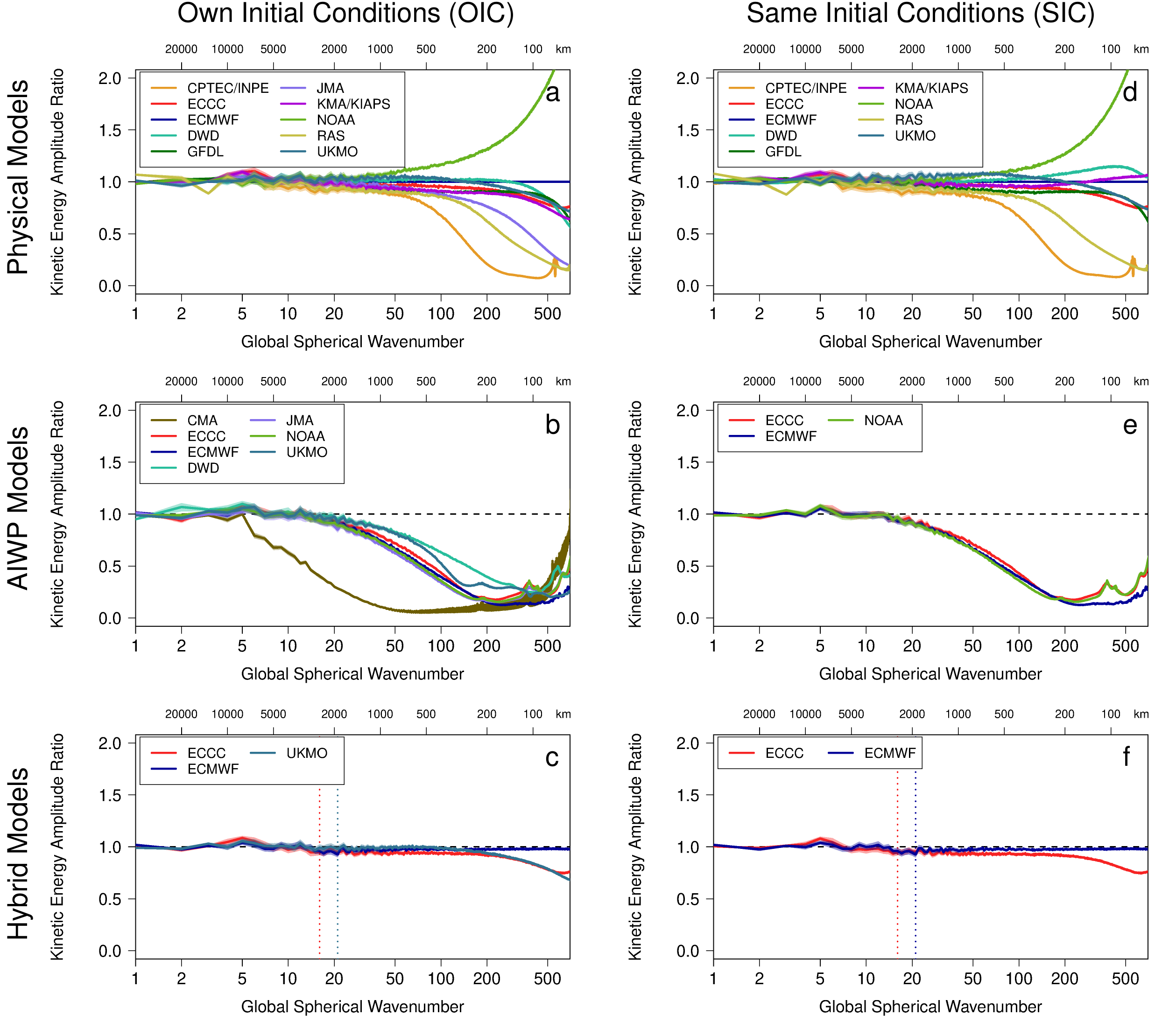}
  \caption{Spectral amplitude ratio for global 250~hPa kinetic energy at day 10 (Eq.~\ref{sar}) \plotlayout.
    A ratio of 1 (unity) is identified with a black dashed line.  Color-coded vertical
    lines for hybrid-model panels (c and f) represent spectral nudging filter cutoffs.}
  \label{spec_rel}
\end{figure*}

Kinetic energy spectra at 250~hPa depict the model state in medium-range
predictions (day 10; Fig.~\ref{spec}).
Also shown are spectral amplitude ratios ($\gamma$), computed following \citet{Husain25} with
respect to the ECMWF analysis as,
\begin{equation}
  \gamma(k) = \sqrt{\frac{1}{N}\sum_{i=1}^N\frac{E_f(i,k)}{E_a(i,k)}} \hspace{0.2in},
  \label{sar}
\end{equation}
for each ($i$) of $N$ predicted ($E_f$) and analyzed ($E_a$) kinetic energies at global spherical
wavenumber $k$. This normalization eases comparison across the many
orders of magnitude of the energy spectrum, with unity providing a reference for
expected amplitude (Fig.~\ref{spec_rel}).

The energy spectra of physical-model forecasts generally follow the expected form (Fig.\ref{spec}a and d),
with slopes of $-3$ that tail off near the effective resolution of the model
\citep{Skamarock04}.  Flattening of the slope below wavenumber
100 ($\sim 400\ \mbox{km}$) in NOAA forecasts appears to represent a transition to 
the mesoscale portion of the energy spectrum \citep{Nastrom85,Stephan22}.
Conversely, damping across a large range of wavenumbers in CPTEC/INPE predictions
is indicative of excess numerical dissipation (Figs.~\ref{spec}a and \ref{spec_rel}a).

All AIWP contributions suffer from significant energetic deficits
at wavenumbers higher than 20 ($\sim 2000\ \mbox{km}$; Figs.~\ref{spec}b and
\ref{spec_rel}b), a well-known consequence of training that minimizes
MSE-based loss functions \citep{Subich25}  for multi-day autoregressive rollouts.
Forecasts from the DWD AIWP model are most active across
a broad range of scales, benefiting primarily from non-rollout training,
supplemented by a nominal grid spacing that is half that of the other AIWP
configurations, and a higher-resolution training dataset.  The
CMA AIWP model is an outlier in the other direction, with significant
energy reductions beginning at planetary scales (wavenumber 5 or
$\sim 8000\ \mbox{km}$).  This is likely the result of a fine-tuning
protocol that includes loss minimization for forecasts out to day-10.
Despite the down-weighting of long-rollout errors, the CMA AIWP forecasts
resemble an ensemble mean
\citep{Bouallegue24}, in terms of both RMSE (Fig.~\ref{rmse}b) and kinetic
energy (Figs.~\ref{spec}b and \ref{spec_rel}b).

Hybrid-model energy spectra generally follow those of their
corresponding physical models (compare Fig.~\ref{spec}a and c,
and Fig.~\ref{spec_rel}a and c).  This means that they predict
significantly more variability across the mesoscale compared to their AIWP
guides (compare Fig.~\ref{spec_rel}b and c), a welcome outcome
from a falsification perspective that increases RMSE
via double penalty as noted at the end of section~\ref{results}\ref{scores}.
Some suppression of kinetic energy occurs near the spectral nudging truncation
length-scales (Fig.~\ref{spec_rel}c), where hybrid-model solutions are nudged
towards energy-deficient AIWP fields (Fig.~\ref{spec_rel}b).
This is a recognized problem that only becomes apparent at longer lead
times \citep{Husain25}; future hybrid systems are
expected to benefit from AIWP guides with more complete power spectra. 

\section{Verification Strategy}
\label{verif}
The WP-MIP verification strategy extends well beyond the basic
evaluations described above. \citet{Casati26} identify nine themes that can be broadly
categorized as focusing on different statistical evaluation archetypes
(section~\ref{verif}\ref{tech}) or atmospheric phenomena (section~\ref{verif}\ref{features}).
Many of the planned verification activities cut across multiple themes,
with coordination through an open task-tracking project \citep{WPMIPgithubverif26}.

\subsection{Statistical Evaluation}
\label{tech}
The development of AI-ready evaluation techniques requires research along
two axes.  The first involves expanding the scope of existing tools and
devising new ways to interpret results.  The second consists of developing
new techniques that are designed to assess characteristics that
are typically taken for granted in physical-model forecasts.

\subsubsection{Traditional Verification}
A baseline of forecast quality needs to be established using the standard
techniques that form the basis of international exchanges \citep{WMO23}.
These include upper-air RMSE, bias, and anomaly correlation, for
which competitive performance should be considered a necessary but not sufficient
condition in the context of AIWP and hybrid models.

\subsubsection{Explainability}
The field of ``explainable AI'' (or ``AI interpretability'') aims to build understanding of AI
behavior.  Although the nonlinearity of atmospheric processes limits the
effectiveness of explainability methods developed in other domains, techniques based
on mechanistic interpretability \citep{Bereska24} have shown promise by identifying the internal representation of
specific weather patterns within AIWP models \citep{MacMillan25,Cheon26}.  In addition to
underpinning trustworthiness, regulatory requirements for
explainability in operational applications are beginning to appear
\citep{Brocker26}.  However, the desire to identify specific model sensitivities is not
unique to AIWP evaluations; it regularly forms the basis for physical model
development as well.  Comparisons of explainability across all model
classes will be possible with WP-MIP forecasts.

\subsubsection{Falsification}
\label{falsification}
Falsification is a way to assess the physical realism
of forecasts: the extent to which predictions represent possible future states of the
atmosphere \citep{Bouallegue26}.  \citet{Moldovan26} show AIWP forecasts with negative precipitation,
a clear example of a physically impossible prediction that reduces the model's
credibility.  More subtle forms of falsification
can be assessed for all classes of models within WP-MIP, including
process-level consistency and energy
budgets whose closure poses a challenge even for established physical
models \citep{Lauritzen22}.

\subsubsection{Spatial Verification}
Spatial verification techniques assess a model's ability to depict realistic
features and patterns by quantifying structure, amplitude, and position
errors \citep{Gilleland10,Dorniger18}.  Blinded evaluations
of WP-MIP precipitation predictions by operational meteorologists will support
the development of verification techniques that quantify perceptions of
forecast utility \citep{Panago24}.

\subsubsection{Prediction of Extremes}
The utility of WP-MIP forecasts for events that lie in the tails of AIWP
training distributions will be assessed using extreme value analyses and
emerging techniques that focus on high-impact weather.  These investigations
of predictions ``when they matter most'' \citep{Panago24} will be supported
by extreme event catalogs that cover the WP-MIP period \citep{Magnusson19b, McGovern25}.

\subsection{Phenomenological Evaluation}
\label{features}
The 1-year WP-MIP period provides a limited sampling of rare or extreme events; however,
atmospheric phenomena can have large societal impacts while being relatively
common on the global domain.  Project forecasts therefore support investigations
of atmospheric features and processes of particular scientific interest or operational
importance.

\subsubsection{Tropical Meteorology}
The improvements in tropical forecast skill achieved by AIWP models are impressive, not
least because these algorithms do not explicitly represent
the convective coupling of tropical wave modes responsible for the majority
of medium-range predictability \citep{Judt20,Keane25}.  This skill extends
to predictions of tropical cyclone track and structural evolution despite
large weak-intensity biases \citep{DeMaria25,Reboita26,Gomez26}, a dichotomy to
be investigated within Subproject~1 (section~\ref{data}\ref{protocol}\ref{sp1}).
The potential for
hybrid models to benefit from the strengths of both physical-model process
representation and AIWP large-scale state guidance for prediction of monsoons
and convectively coupled waves will also be explored using WP-MIP forecasts.

\subsubsection{Subseasonal-to-Seasonal}
Forecast contributions to Subproject~2
(section~\ref{data}\ref{protocol}\ref{sp2}) will permit preliminary
AIWP and hybrid-model week-2 forecast skill assessments.  Although the WP-MIP protocol lacks ingredients
required for robust S2S evaluation (ensembles, monthly lead
times, and sampling across different background states),
predictions of relevant phenomena, including the MJO, NAO, the Southern Annular
Mode, and tropical-extratropical teleconnections will be studied as prerequisites for
potential extended-range forecast skill.

\subsubsection{Case and Feature-Based Studies}
The 6-hourly WP-MIP forecast fields (section~\ref{data}\ref{protocol})
are designed to support both individual case studies and
broader assessments of specific phenomena.  The latter will involve development
and refinement of AI-adapted feature identification and tracking algorithms for
application to studies of atmospheric rivers, midlatitude cyclones, fronts, and
other atmospheric phenomena \citep[e.g., ][]{Zhang25}.

\subsubsection{Regional Studies}
The global WP-MIP forecast domain enables locally focused studies of each
model's ability to generate skillful predictions of regionally relevant features.
With project members from all WMO regions, such studies will help to
increase local expertise while providing actionable information about predictive skill.
Both of these outcomes contribute directly to the broader objectives of the WMO EW4All
initiative.

\section{WP-MIP and Beyond}
\label{discussion}
The WMO Weather Prediction Model Intercomparison Project (WP-MIP) has assembled
an extensive archive of forecasts from physically based, AIWP, and hybrid
models being developed at national meteorological and hydrological services around
the world.  This dataset has enabled project verification teams to begin
developing ``AI-ready'' evaluation techniques that are capable of quantifying
the strengths and weaknesses of the different modeling paradigms.

Future iterations of WP-MIP or similar projects will be needed to ensure
that findings related to relative forecast skill and verification best-practices
remain current in the rapidly evolving NWP landscape.  Updated AIWP training
regimens will overlap with the 2024 period used here, particularly as model
developers explore alternative fine-tuning techniques.  Such practices may necessitate
changes to the traditional hindcast-based experimental protocol adopted for
WP-MIP, although the design of such a replacement remains to be determined.

One important element that is missing from the WP-MIP archive is
ensemble-based prediction.  Recent AIWP successes in this domain suggest
that any future intercomparison project will need to include a probabilistic
component.  Such an expansion will also facilitate research into extremes
and enable more robust assessments of the potential for improved S2S
predictions.  Updates to TIGGE may satisfy many of the requirements for a future WP-MIP iteration;
however, the evolution of systems contributing to TIGGE makes well-controlled
intercomparison difficult. Discussions with the coordinators of the
WMO Bridging the Gap project \citep{Pic26} are underway to design a protocol that
limits data volume while enabling robust ensemble intercomparison.

By facilitating the assessment of physically based, AIWP, and hybrid models being
developed at operational forecasting centers, and accelerating the development of AI-ready
verification techniques, WP-MIP represents an important WMO contribution to
global NWP research.  The lessons learned from this project will inform
decisions about the deployment of the next generation of operational models.
With their ability to generate high-quality predictions in resource-limited
environments, the AIWP components of such systems have the potential to
expand the availability of critical weather information, a key EW4All objective
that promises to protect lives and property around the globe.

\clearpage
\acknowledgments
The authors thank editor Craig Schwartz and all reviewers (including
Brian Henn and Jacob Radford) for the insightful and constructive
suggestions that helped to prepare this study for publication.
This work is a contribution to Earth System Modelling and Observations (ESMO), a core
project of the World Climate Research Programme (WCRP) of the World Meteorological Organization
(WMO), which advances coordination and innovation in Earth system modeling and observational
activities. We acknowledge the WCRP for coordinating this international research effort and
the German Federal Ministry of Research, Technology and Space (BMFTR) for funding through the
ESMO IPO project (Grant No. 01LP2312A).
A. McGovern is supported by the US National Science Foundation (2019758). 
G. Skok is supported by the Slovenian Research and Innovation Agency (Research Core Funding P1-0188).
M. Tolstykh is supported by the Russian Science Foundation (25-17-00314).
Z. Wang is supported by the Office of Naval Research (Grant N000141812216).
G. Zhang is supported by the US National Science Foundation (RISE-253055).

%
%
\datastatement
All data used in this study were retrieved from the WP-MIP forecast portal \citep{WPMIPportal26}.  The
recommended citation for such data is, ``the data from [CONTRIBUTOR LIST] used
in this work were distributed by the Weather Prediction Model Intercomparison Project
(https://www.wcrp-esmo.org/activities/wp-mip) under the CC-BY-4.0 license.''  The
software used for analysis and plotting is available at https://github.com/WP-MIP/verification.


%






%



\bibliographystyle{ametsocV6}
\bibliography{refer}

\end{document}